\documentstyle[prl,aps,twocolumn,floats,epsf]{revtex}
\begin{document}
\draft
\twocolumn[\hsize\textwidth\columnwidth\hsize\csname @twocolumnfalse\endcsname
\title{Distorted Perovskite with $e_g^1$ Configuration as a Frustrated Spin System}
\author{T.\ Kimura,$^1$ S.\ Ishihara,$^2$ H.\ Shintani,$^1$ 
T.\ Arima,$^3$ K.\ T.\ Takahashi,$^1$ K.\ Ishizaka,$^1$ and Y.\ Tokura$^1$}
\address{$^1$Department of Applied Physics, University of Tokyo, Tokyo 113-8656, Japan}
\address{$^2$Department of Physics, Tohoku University, Sendai 980-8578, Japan}
\address{$^3$Institute of Materials Science, University of Tsukuba, Tsukuba 305-8573, Japan} 
\date{\today}
\maketitle
\begin{abstract}
The evolution of spin- and orbital-ordered states has been investigated for a series of insulating perovskites $R$MnO$_3$ ($R$=La,Pr,Nd,...). 
$R$MnO$_3$ with a large GdFeO$_3$-type distortion is regarded as a frustrated spin system having ferromagnetic nearest-neighbor and antiferromagnetic (AF) next-nearest-neighbor (NNN) interactions within a MnO$_2$ plane. 
The staggered orbital order associated with the GdFeO$_3$-type distortion induces the anisotropic NNN interaction, and yields unique {\it sinusoidal} and {\it up-up-down-down} AF ordered states in the distorted perovskites with $e_g^1$ configuration. 
\end{abstract} 
]
\narrowtext 

Common electronic characteristics exists in perovskite manganites 
$R$MnO$_3$ and nickelates $R$NiO$_3$ ($R$=trivalent lanthanoids). 
On the Mn$^{3+}$ and Ni$^{3+}$ sites with the $e_g^1 t_{2g}^3$ 
and $e_g^1 t_{2g}^6$ configurations, respectively, 
the $e_g$ orbital is doubly  degenerate and 
the $t_{2g}$ orbital degree of freedom is quenched. 
It is widely recognized that the layered-type ($A$-type) antiferromagnetic (AF) structure 
in LaMnO$_3$ is understood from the view point of the anisotropic superexchange (SE) 
interaction under the directional order of orbital \cite{KIKugel1,SIshihara1}. 
On the other hand, the spin structure in nickelates ($R$$\neq$La) is distinct from the $A$-type AF; 
the so-called "up-up-down-down"-type one where 
two Ni sites of "up" spins are followed by two sites of "down" spins 
along the principal axes in the cubic unit cell.
Origin of this unusual magnetic order has been a long-standing 
question, 
as well as its relations to metal-insulator transition, orbital order (OO), and charge disproportionation \cite{JLGarcia1,JAAlonso1,TMizokawa1}. 
Recently, a similar spin structure, {\it i.e.} 
the up-up-down-down order in a MnO$_2$ plane 
($E$-type AF order in the Wollan-Koehler notation \cite{EOWollan1}),  
is found in a manganite, HoMnO$_3$ \cite{AMunoz2} with a significantly distorted perovskite structure. 
This has to be a bridge between the well-understood $A$-type AF in manganites and 
the unique magnetic ground state in nickelates. 

\begin{figure}
\epsfxsize=0.65\columnwidth
\centerline{\epsffile{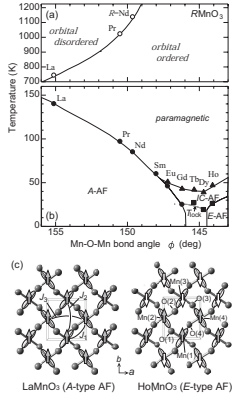}}
\caption{
Orbital (a) and spin (b) ordering temperatures of $R$MnO$_3$ 
as a function of the inplane Mn-O-Mn bond angle. 
 (c) Crystal structures of LaMnO$_3$ and HoMnO$_3$. 
Spin (arrows) and orbital (lobes) ordered features are also illustrated. 
The stack of spin and orbital order along the $c$-axis is staggered and 
uniform order, respectively, for the both compounds. 
}
\label{fig1}
\end{figure}

In this Letter, 
we examine systematically the magnetic and orbital structures 
in a series of $R$MnO$_3$ as a function of the ionic radius ($r_R$) of $R$. 
The most significant effect on the crystal structure by decreasing $r_R$ is an enhancement of the cooperative rotation of the MnO$_6$ octahedra
(the GdFeO$_3$-type distortion) characterized by the decrease of Mn-O-Mn bond angle $\phi$. 
Let us first summarize in Fig.\ \ref{fig1} the orbital (a) and spin (b) ordering temperatures ($T_{\rm OO}$ and $T_{\rm N}$, respectively) on Mn sites of $R$MnO$_3$ as a function of $\phi$, which is based on both the present and former studies \cite{AMunoz2,ZJirak1,SYWu1,SQuezel1}. 
Here, we adopt the $\phi$ at room temperature \cite{JAAlonso2}. 
The $T_{\rm OO}$ monotonically increases with decreasing $r_R$, while 
the magnetic transition occurs from the $A$-type AF to the $E$-type one through the incommensurate structure. 
We argue that the combination of OO and next-nearest-neighbor (NNN)
SE interaction brings about a nontrivial effect on the magnetic ground state 
in the systems with the orbital degeneracy and 
the large GdFeO$_3$-type distortion. 
Microscopic calculation shows that the magnetism in this system is mapped onto the frustrated 
spin model which well reproduces the phase diagram of $R$MnO$_3$. 

A series of $R$MnO$_3$ ($R$=La$-$Dy) crystals were grown by the floating zone method. 
We made powder x-ray diffraction measurements on the obtained crystals at room temperature, and confirmed that all the crystals show the $Pbnm$ orthorhombic structure. 
Magnetization at 0.5 T was measured with a SQUID magnetometer. 
Specific heat was measured using a relaxation technique. 
Resistivity measurements were made by a standard 4-probe method in a flow of Ar gas up to $\sim$1200 K. 

\begin{figure}
\epsfxsize=0.65\columnwidth
\centerline{\epsffile{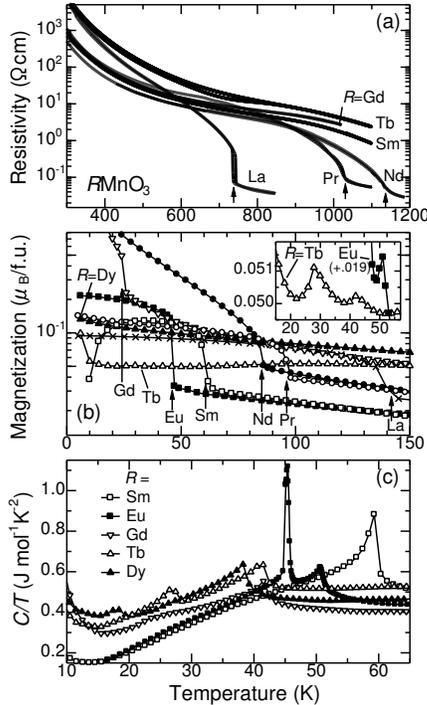}}
\caption{
Temperature profiles of (a) resistivity, (b) magnetization $M$, and (c) specific heat divided by 
temperature $C/T$ for $R$MnO$_3$ crystals. 
Vertical arrows in (a) and (b) indicate $T_{\rm OO}$ and $T_{\rm N}$ for the Mn moment, respectively. 
The inset magnifies the $M$ of $R$=Tb and Eu in the vicinity of $T_{\rm N}$. 
}
\label{fig2}
\end{figure}

Let us show in Fig.\ \ref{fig2} the experimental data [temperature ($T$) profiles of (a) resistivity $\rho$,  (b) magnetization $M$, and (c) specific heat divided by temperature $C/T$] which compose the phase diagram of Figs.\ \ref{fig1}(a) and \ref{fig1}(b). 
As shown in Fig.\ \ref{fig2}(a), all the crystals show insulating behaviors over the whole $T$ range investigated here. 
For a LaMnO$_3$ crystal, the $\rho$ shows an abrupt drop toward high $T$ at $T_{\rm OO}$$\approx$747 K in accord with the cooperative Jahn-Teller (JT) or orbital-ordering transition \cite{JRodriguez1}. 
With decreasing $r_R$, the anomaly in $\rho$, i.e.\ $T_{\rm OO}$, is shifted toward higher $T$ (indicated by arrows). 
In the crystals with smaller $r_R$ than $r_{\rm Nd}$, however, 
no such anomaly was observed up to 1500 K. 
Thus, the OO state associated with the cooperative JT distortion is extremely stable 
in $R$MnO$_3$ with small $r_R$. 

As shown in Fig.\ \ref{fig2}(b), a steep rise of $M$ toward lower $T$ (indicated by vertical arrows) is observed in most of the crystals. 
The anomaly in $R$=La$-$Nd crystals well corresponds to 
$T_{\rm N}$ for the $A$-type AF order \cite{EOWollan1,ZJirak1,SYWu1}. 
The $T_{\rm N}$ falls monotonically with decreasing $r_R$ from La to Gd. 
A similar jump of $M$ attributed to the spin ordering of Mn site is not observed in $R$=Tb and Dy crystals with small $r_R$. (The anomaly in $M$ below 10 K is related to the ordering of $R$-site $f$-moment.) 
In TbMnO$_3$, however, the $M$ exhibits two sharp peaks at $\sim$42 K and $\sim$27 K [the inset of Fig.\ \ref{fig2}(b)]. 
Figure \ref{fig2}(c) shows the $C/T$ for crystals with smaller $r_R$. 
In SmMnO$_3$, the jump of $C/T$ at $\sim$59 K nicely agrees with the steep rise of $M$, and can be assigned to the $A$-type AF ordering. 
A remarkable feature in the $C/T$ of EuMnO$_3$ is the sharp peak at $\sim$46 K, as well as the jump at $\sim$51 K. The sharp 46K-peak is suggestive of the first order phase transition. 
In crystals with smaller $r_R$ ($R$=Gd$-$Dy), a rather broader peak is observed at $\sim$40 K. 
In addition, another broad peak feature is evident in the $T$ region 18-26 K. 
Among them, the $T$ evolution of spin structure has been investigated for TbMnO$_3$ by neutron diffraction measurements \cite{SQuezel1}. 
The observed peaks in $C$$/$$T$ and $M$ at $\sim$42 K for TbMnO$_3$ correspond to the onset of the sine-wave ordering of the Mn moments with the wave vector of $(0,k_s,0)$. 
The $k_s$ ($\sim$0.295) at $T_{\rm N}$ is incommensurate (IC) and decreases with decreasing $T$, and becomes nearly constant ($k_s$=0.28) below $\sim$30 K. 
The anomalies in $C/T$ and $M$ at $\sim$27 K are in good agreement with the $T$ where $k_s$ is locked at a constant value ($T_{\rm lock}$). 
With further decreasing $r_R$, 
Mu\~noz {\it et al.} \cite{AMunoz2} reported that in polycrystalline  HoMnO$_3$ ($T_{\rm N}$=41 K) 
the IC$-$to$-$commensurate (CM) magnetic phase transition takes place 
at $\sim$26 K, 
where the wave vector is $(0,k_s,0)$ [0.4$\leq$$k_s$$<$0.5  ($T$-dependent) for the IC phase 
and $k_s$$=$$\frac{1}{2}$ for the CM one]. 

As displayed in Fig.\ \ref{fig1}, $T_{\rm OO}$ steeply increases with decreasing $\phi$, whereas $T_{\rm N}$ for the $A$-type AF order monotonically decreases. 
With the suppression of the $A$-type AF order, the IC sinusoidal magnetic structure which propagates along the $b$-axis appears. 
With further decreasing $\phi$, the CM magnetic structure with the wave vector of $(0,\frac{1}{2},0)$ turns up at the ground state in HoMnO$_3$ \cite{AMunoz2}. 
The CM magnetic structure can be identified with 
the "up-up-down-down" spin structure within the $ab$-plane or 
the $E$-type AF structure. 
To visualize the modification of the crystallographic and magnetic structures by the decrease of $\phi$, we illustrate in Fig.\ 
\ref{fig1}(c) the projection of the fundamental crystal structure of LaMnO$_3$ and HoMnO$_3$ along the $c$-axis. 

In $R$MnO$_3$ with a small GdFeO$_3$-type distortion, 
such as $\rm LaMnO_3$, 
the staggered [$d_{3x^2-r^2}$$/$$d_{3y^2-r^2}$]-type OO 
is responsible for the $A$-type AF order. 
There are the ferromagnetic (FM) SE interaction between nearest-neighbor (NN) $e_g$ spins 
and the AF one ($J^{t_{2g}}_{AF}$) between NN $t_{2g}$ spins. 
The latter is superior along the $c$-axis \cite{SIshihara1}.
In $R$MnO$_3$ with significant GdFeO$_3$-type distortion (small $r_R$), 
the FM SE interaction is weakened due to 
reduction of the transfer intensity of an $e_g$ electron. 
However, such an argument based on the 
NN interactions is not enough 
to explain the $E$-type AF or sinusoidal magnetic order; 
the inversion symmetry of the spin 
alignment is broken in the $ab$ plane for the $E$-type AF structure, 
in spite that the two NN bonds along the opposite directions 
are equivalent from the crystal structural point of view. 

The crucial effect caused by the significant GdFeO$_3$-type distortion  
is the SE interaction between NNN Mn sites. 
It is evident in Fig.\ \ref{fig1}(c) that the enhancement of the 
GdFeO$_3$-type distortion shortens the distance between O(2) and O(4)  
[e.g.\ the O(2)-O(4) length is $\approx$3.4 ${\rm \AA}$ for LaMnO$_3$ 
and $\approx$3.0 ${\rm \AA}$ for HoMnO$_3$ at room temperature \cite{JAAlonso2}].
This shortening enhances the SE 
interaction between NNN sites through 
Mn-O(2)-O(4)-Mn exchange paths.  
Under the staggered OO, in addition to the GdFeO$_3$-type distortion, 
the two NNN SE interactions along the different directions become inequivalent; 
the interaction between Mn(1) and Mn(3) (along the $b$-axis) 
is stronger than that between Mn(2) and Mn(4) (along the $a$-axis).
Since the occupied orbitals in Mn(1) and Mn(3) are the same, this SE interaction 
is the AF one which brings about the spin frustration. 

We present the theoretical prescription for 
the combination effect of the GdFeO$_3$-type distortion 
and the staggered OO. 
The Hamiltonian adopted here is 
the spin-orbital model which is known to describe well  
the orbitally degenerate manganites \cite{SIshihara1}; 
${\cal H}$$=$${\cal H}_{J}$$+$${\cal H}_{H}$$+$${\cal H}_{AF}$. 
The main term ${\cal H}_J$ is 
the exchange interaction between intersite 
$e_g$ spins and orbitals schematically written as 
${\cal H}_J$$=$$\sum_m J_m \sum_{\langle i j \rangle}(a_m \vec S_i \cdot \vec S_j$$+$$b_m)h_m(\vec T_i, \vec T_j)$. 
$m$ is an index classifying the exchange processes, 
$J_m$ indicates the SE interactions, $a_m$ and $b_m$ are the constants, and 
$\vec S_i$ is the spin operator of the  $e_g$ electron. 
The orbital part $h_m(\vec T_i, \vec T_j)$ is represented
by the pseudospin operator  
$\vec T_i$ with a magnitude 1/2. 
${\cal H}_{H}$ and ${\cal H}_{AF}$ in ${\cal H}$
are the Hund coupling between $e_g$ and 
$t_{2g}$ spins, and the AF interaction $J_{AF}^{t_{2g}}$ 
between NN $t_{2g}$ spins, respectively.  
Beyond the conventional spin-orbital model, 
the SE interactions between NNN Mn sites are considered in ${\cal H}_{J}$. 
The effective electron transfer 
$t_{ij}^{\gamma \gamma'}$ between $i$ and $j$ Mn sites with  
$\gamma$ and $\gamma'($$=$$3z^2$$-$$r^2, x^2$$-$$y^2)$ orbitals   
occurs through the O $2p$ orbitals \cite{transfer}. 
For example, for the Mn(1)-Mn(3) pair [see Fig.~\ref{fig1}(c)], 
possible exchange paths are [Mn(1)$-${O(1),O(4)}$-${O(2),O(3)}$-$Mn(3)]. 
Both the GdFeO$_3$-type and JT-type distortions 
are introduced in $t_{ij}^{\gamma \gamma'}$  
through the Slater-Koster formulae \cite{JCSlater}. 

\begin{figure}
\epsfxsize=0.65\columnwidth
\centerline{\epsffile{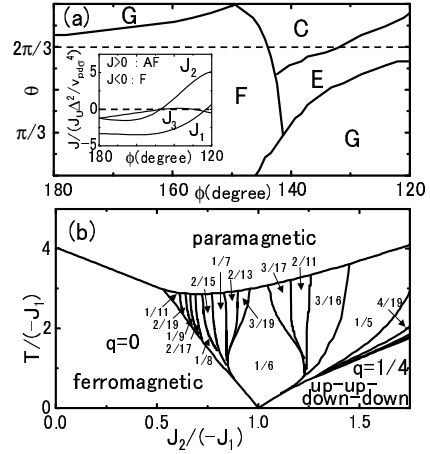}}
\caption{
(a) Magnetic phase diagram at $T$$=$$0$. 
$\theta$ is the orbital mixing angle. 
The broken line indicates 
the $[d_{3x^2-r^2}/d_{3y^2-r^2}]$-type OO.  
$C$, $E$, $G$ and $F$ imply the $C$-, $E$-,  
$G$-type AF and FM phases, respectively \protect\cite{mean}. 
The inset shows the effective SE interactions between NN Mn spins ($J_1$), 
NNN Mn spins along the $b$ axis ($J_2$), 
and NNN Mn spins along the $a$ axis ($J_3$) at $\theta=2\pi/3$ \protect\cite{parameter}. 
(b) Mean-field magnetic phase diagram of the 2D $J_1-J_2-J_3$ model 
with $J_3/J_1=0.01$. 
The each phase is characterized by 
a wave number $q=M/N$ of the spin structure 
along the $b$ axis. $q=0$ and $1/4$ correspond to 
the $A$- and $E$-type AF states in $R$MnO$_3$, respectively.  
}
\label{fig3}
\end{figure}

The magnetic phase diagram  
is calculated by the mean field approximation at $T$$=$$0$ [Fig.~\ref{fig3}(a)] \cite{parameter,mean} 
in the two-dimensional (2D) square lattice,  
since the AF spin alignment along the $c$-axis due to $J_{AF}^{t_{2g}}$ remains 
unchanged in a series of $R$MnO$_3$. 
The staggered OO with two sublattices  
is of the $[\theta/-\theta]$-type characterized by 
the mixing angle:  
$|\theta \rangle =\cos (\frac{\theta}{2}) |d_{3z^2-r^2} \rangle 
+ \sin(\frac{\theta}{2}) |d_{x^2-y^2}\rangle $. 
The $[d_{3x^2-r^2}/d_{3y^2-r^2}]$-type OO corresponds to 
$\theta$$=$$2\pi/3$. 
Without the GdFeO$_3$-type distortion, 
the FM order in the $ab$-plane, 
corresponding to the $A$-type AF order in the three dimensional lattice, 
appears for $\theta$$<$$1.75\pi$. 
With decreasing $\phi$, 
the $E$-type AF phase of the present interest appears for $1.75 \pi$$<$$\theta$$<$$2.5 \pi$ and $\phi$$<$$143^\circ$. 
This result agrees semiquantitatively with  
the experiments. 
The remarkable change 
with decreasing $\phi$ 
is seen in the SE interaction between Mn(1) and Mn(3) along the $b$-axis;  
it turns to a strong AF interaction from a weak FM one 
[see $J_2$ in the inset of Fig.~\ref{fig3}(a)] 
as well as weakening of the NN 
FM one ($J_1$) \cite{LEGontchar,THotta1}. 

The essence of magnetic properties in this system is mapped onto 
the 2D frustrated Heisenberg model for $S$$=$$2$ with 
FM NN interaction ($J_1$), 
AF NNN one along the $b$ axis ($J_2$), 
and weak FM NNN along the $a$ axis ($J_3$).  
The finite $T$ phase diagram is obtained by the mean field approximation [Fig.~\ref{fig3}(b)]. 
A periodicity, $N$, of the spin structure is taken up to 20 
along the $a$, $b$ and $a$$\pm$$b$ directions, 
and each phase is characterized by the wave number $q$$=$$M/N$. 
The phase diagram shows a similar 
topological structure to that in the ANNNI model \cite{PBak1,PBak2}; 
numerous long-range orders between the FM $(q$$=$$0)$ 
and "up-up-down-down"-type AF ($q$$=$$1/4$) phases, 
that is the so-called Devil's flower. 
The calculated results qualitatively 
reproduce the phase diagram of $R$MnO$_3$ in Fig.~\ref{fig1}(b). 
(Note that the $A$-type AF state is regarded as the 2D FM state.) 

\begin{figure}
\epsfxsize=0.8\columnwidth
\centerline{\epsffile{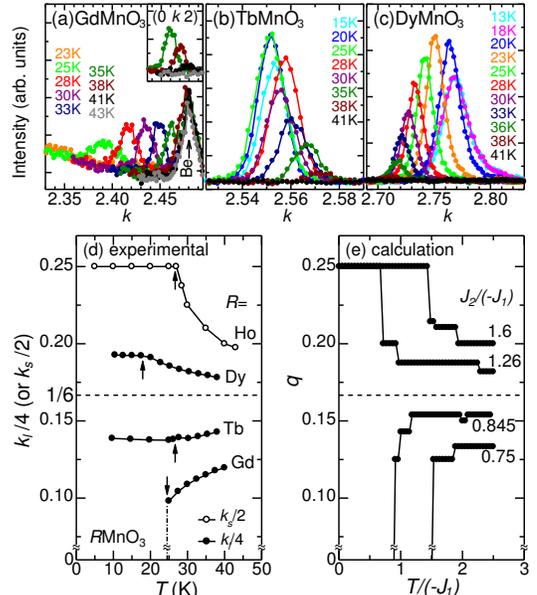}}
\caption{(Color) 
(a)-(c): X-ray diffraction $k$-scans along (0,$k$,3) at various $T$ for $R$=Gd, Tb, and Dy crystals, measured at BL-4C of PF-KEK, Tsukuba. 
$T$-profiles of the wave number of modulated magnetic structure obtained by  (d) experiment and (e) calculation. The arrows denote $T_{\rm lock}$. The data of $k_s$ in HoMnO$_3$ were taken from ref.\ \protect\cite{AMunoz2}.  
}
\label{fig4}
\end{figure}

Further supporting evidence is needed to confirm the validity of the present scenario. 
However, it is difficult to investigate the spin structure by the neutron diffraction for compounds with Gd and Dy elements because of their large neutron scattering cross sections. 
Hence, we overcome the problem by measurements of single crystal x-ray diffraction. 
Figures \ref{fig4}(a)- \ref{fig4}(c) show x-ray diffraction scans along (0,$k$,3) at various $T$ for $R$=Gd, Tb, and Dy crystals \cite{xray}. 
For all the crystals, additional superlattice peaks appear at the wave vector (0,$k_l$,$l$) for integer $l$ below $T_{\rm N}$. 
In TbMnO$_3$, the $k_l$ is $\sim$0.57 at $T_{\rm N}$$\sim$40 K, 
decreases with decreasing $T$, and becomes nearly constant ($k_ l$$\sim$0.55) below $T_{\rm lock}$$\sim$27 K. 
The value of $T$-dependent $k_l$ is almost twice as large as that of $k_s$. 
It is well-known that the crystallographic deformations at magnetic ordering are due to the {\it exchange striction} \cite{JSSmart1}. 
The observed superlattice reflections due to the atomic displacement can be regarded as the second harmonic peaks magnetoelastically induced by sinusoidal AF order. 
Hence, a half value of $k_l$ could represent $k_s$. 
The $T$-profiles of the wave number $k_s$=$k_l/2$ obtained by experiments are compared with those calculated for the representative values of $J_2/(-J_1)$ [Figs.\ \ref{fig4} (d) and \ref{fig4}(e)]. 
The theoretical results are in quantitatively good agreement with experiments 
in terms of $T$- and $R$-dependence, 
which strongly suggests that the present modeling approach is proper for understanding the phase diagram of $R$MnO$_3$. 

We examined the evolution of magnetic and orbital states 
in a series of $R$MnO$_3$ as a function of the ionic radii $r_R$ in $R$. 
The $T_{\rm N}$ of the $A$-type AF order steeply decreases with the decrease of $r_R$. 
Eventually the up-up-down-down type ($E$-type) AF order 
appears in $R$=Ho via the sinusoidal magnetic order in 
$R$=Tb. 
Such curious AF ordered states in $R$MnO$_3$ 
can be explained in a scenario of the spin frustration caused by 
the combination of the significant GdFeO$_3$ distortion 
and the staggered OO; 
the former enhances the NNN SE interaction, and the latter causes the anisotropy in the NNN SE interaction. 
This scenario can be also applicable to the up-up-down-down AF order observed in $R$NiO$_3$ with the distorted perovskite structure. 

We thank T.\ Hotta, D.\ I.\ Khomskii, E.\ Dagotto, T.\ Mizokawa, and 
N.\ Nakamura for helpful discussions. 
This work was supported by KAKENHI from MEXT, Japan.

\begin{references}
\bibitem{KIKugel1}
K.\ I.\ Kugel and D.\ I.\ Khomskii, Sov.\ Phys.\ JETP {\bf 37}, 725 (1973); 
Sov.\ Phys.\ Usp.\ {\bf 25}, 232 (1982). 

\bibitem{SIshihara1}
S.\ Ishihara {\it et al.}, 
Phys.\ Rev.\ B {\bf 55}, 8280 (1997). 

\bibitem{JLGarcia1}
J.\ L.\ Garc\'ia-Mu\~noz {\it et al.}, Phy.\ Rev.\ B {\bf 50} 978 (1994). 

\bibitem{JAAlonso1}
J.\ A.\ Alonso {\it et al.}, Phys.\ Rev.\ Lett.\ {\bf 82}, 3871 (1999). 

\bibitem{TMizokawa1}
T.\ Mizokawa {\it et al.}, Phy.\ Rev.\ B {\bf 61}, 11263 (2000). 

\bibitem{EOWollan1}
E.\ O.\ Wollan and W.\ C.\ Koehler, Phys.\ Rev.\ {\bf 100}, 545 (1955). 

\bibitem{AMunoz2}
A.\ Mu\~noz {\it et al.}, 
Inorg.\ Chem.\ {\bf 40}, 1020 (2001). 

\bibitem{ZJirak1}
Z.\ Jirak {\it et al.}, J.\ Magn.\ Magn.\ Mater.\ {\bf 53}, 153 (1985). 

\bibitem{SYWu1}
S.\ Y.\ Wu {\it et al.}, J.\ Appl.\ Phys.\ {\bf 87}, 5822 (2000). 

\bibitem{SQuezel1}
S.\ Quezel {\it et al.}, Physica B {\bf 86-88}, 916 (1977). 

\bibitem{JAAlonso2}
J.\ A.\ Alonso {\it et al.}, Inorg.\ Chem.\ {\bf 39}, 917 (2000). 
(We use $\phi$ of $R$=Sm, Eu, and Gd samples by interpolating from a nearly linear relation between $\phi$ and $r_R$.) 

\bibitem{JRodriguez1}
J.\ Rodr\'guez-Carvajal {\it et al.}, 
Phys.\ Rev.\ B {\bf 57}, R3189 (1998). 

\bibitem{transfer}
$t_{ij}^{\gamma \gamma'}$ between NN and NNN sites are given by 
$\frac{1}{\Delta}\sum_{k \alpha} t_{k i}^{p_\alpha d_\gamma}t_{k j}^{p_\alpha d_{\gamma'}}$ 
and 
$\frac{1}{\Delta^2}
\sum_{k l \alpha \beta} t_{k i}^{p_\alpha d_{\gamma}} t_{k l}^{p_\alpha p_\beta} 
t_{l j}^{p_\beta d_{\gamma'}}$, 
respectively, with the charge transfer energy $\Delta$, 
and  the transfer integral between Mn$3d$ (O$2p$) and O$2p$ ions  
$t_{k i}^{p_\alpha d_\gamma}$ ($t_{k l}^{p_\alpha p_{\alpha'}}$). 

\bibitem{JCSlater}
J.\ C.\ Slater and G.\ F.\ Koster, Phys.\ Rev. {\bf 94}, 1498 (1954). 

\bibitem{parameter}
We chose 
the charge transfer energy $\Delta$$=$2eV, 
the transfer integral between $3d$ and $2p$ orbitals $v_{pd \sigma}$$=$$-1.58$eV, 
that between $2p$ orbitals $v_{pp \sigma}$$=$$0.5$eV, $J_{U'-I}/J_U$$=$$3$ and $J_{AF}^{t_{2g}}/J_U$$=$$0.01$, 
where $J_{U'-I}$ and $J_{U}$ are the SE interactions in ${\cal H}_J$ \cite{SIshihara1}. 

\bibitem{mean}
Six spin structures are assumed in the 2D lattice: 
the FM structure ($F$), 
the checker board-type AF one ($G$), 
the chain-type AF ones along the $a-b$ axis ($C$) and that along the $a+b$ axis ($C'$), 
and 
the up-up-down-down-type AF ones with the zigzag chains along the $a$ axis ($E$) 
and that along the $b$ axis ($E'$). 

\bibitem{LEGontchar}
L.\ E.\ Gontchar and A.\ E.\ Nikiforov, Phys.\ Rev.\ B {\bf 66}, 014437 (2002). 

\bibitem{THotta1}
T.\ Hotta {\it el al.} also study the $E$-type AF order by using another model (cond-mat/0211049). 

\bibitem{PBak1}
P.\ Bak and J.\ von Boehm, Phys. Rev. B {\bf 21}, 5297 (1980). 

\bibitem{PBak2}
P.\ Bak, Rep.\ Prog.\ Phys. {\bf 45}, 587 (1982). 

\bibitem{xray}
Since the diffraction by a Be window overlaps that of the superlattice at (0,$k$,3)  near $T_{\rm N}$ for GdMnO$_3$, we show in the inset of Fig.\ 4(a) (0,$k$,2) scans near $T_{\rm N}$. 

\bibitem{JSSmart1}
J.\ S.\ Smart and S.\ Greenwald, Phys.\ Rev.\ {\bf 82}, 113 (1951). 

\end{references}
\end{document}